\documentclass[a4paper,10pt]{revtex4}
\usepackage{mathrsfs}
\usepackage{graphicx}
\usepackage{latexsym}
\usepackage{amsmath}
\usepackage{amssymb}
\usepackage{textcomp}
\usepackage{amsbsy}
\usepackage{graphics}
\usepackage{epstopdf}
\usepackage{color}

\begin{document}

\title{Cosmological quantum entanglement : A possible testbed for the existence of Kalb-Ramond field}

\author{Tanmoy~Paul,$^{1,2}$\,\thanks{pul.tnmy9@gmail.com}
Narayan~Banerjee$^{3}$\,\thanks{narayan@iiserkol.ac.in}} \affiliation{ $^{1)}$ Department of Physics, Chandernagore College, Hooghly - 712 136.\\
$^{2)}$ Department of Theoretical Physics,\\
Indian Association for the Cultivation of Science,\\
2A $\&$ 2B Raja S.C. Mullick Road,\\
Kolkata - 700 032, India.\\
$^{3)}$ Department of Physical Sciences, Indian Institute of Science Education and Research Kolkata,
Mohanpur Campus, Nadia, West Bengal 741246, India. }

\tolerance=5000

\begin{abstract}
In the present paper, we explore the possible effects of a second rank antisymmetric tensor field, known as Kalb-Ramond (KR) field, on cosmological 
particle production as well as on quantum entanglement for a massive scalar field propagating in a four dimensional FRW spacetime evolves 
through a symmetric bounce. For this 
purpose, the scalar field is considered to be coupled with the KR field and also with the Ricci scalar via the term $\sim \xi R\Phi^2$ (with 
$\xi$ be the coupling). The presence of KR field spoils the conformal symmetry of a massless scalar field even for $\xi = 1/6$ in four dimensional 
context, which has interesting consequences on particle production and consequently on quantum entanglement as we will discuss. 
In particular, the 
presence of KR field in a FRW bouncing universe allows a greater particle production and consequently the upper bound of the entanglement entropy 
becomes larger in comparison to the case when the KR field is absent. This may provide an interesting testbed for the existence 
of Kalb-Ramond field in our universe.
\end{abstract}

\maketitle

\section{Introduction}
The inflationary paradigm \cite{guth,starobinsky,linde1,linde2,rubakov,lyth} is one of the two most successful scenarios that can 
consistently describe the primordial era of our Universe, while the second scenario being that of 
the bounce cosmology\cite{brandenberger}. Both scenarios can predict a nearly scale invariant 
power spectrum and a small amount of gravitational radiation, which are also verified and 
tightly constrained by the latest Planck \cite{planck}. 
Despite the enormous successes, a consistent cosmological model 
is still riddled with some questions like (1) the late time acceleration of our universe \cite{acc1,acc2} implying that most of the content of the universe 
is some form of dark energy whose nature remains unknown to us, (2) the mystery of dark matter remains elusive inspite of the latest progress 
of the Large Hadron Collider in the context of elementary Particle Physics \cite{aad1,aad2,vk1,vk2}.\\

Another important ingredient for a cosmological model (more generally in the context of gravity) is missing: we do not fully understand 
the nature of quantum gravity responsible for early universe phenomena like inflation or bouncing. Without a generally accepted quantum 
theory of gravity where the backreaction of quantum fields on the curvature of spacetime should be taken into account, the quantum field 
theory in a curved spacetime without the backreaction is normally considered so far \cite{davies}. 
In this regard, the dynamics of the background spacetime 
has non-trivial effects on quantum fields propagating on that spacetime when compared with their flat spacetime counterparts. In particular, the 
interaction of a quantum scalar field with a time dependent classical FRW spacetime excites a definite number 
of scalar particles from an infinite-past-vacuum-state and consequently the particles become quantum entangled in the asymptotic future 
\cite{ball,fuentes,steeg,genovese,martinez1,martinez2,martinez3,pierini1,pierini2,pierini3,nakai}. Such 
entanglement may be quantified by von-Neumann or Renyi entropy \cite{nakai}, 
however in the present paper we stick to the von-Neumann entropy. Moreover it is shown 
that such quantum correlations may be present today as a remnant of the primitive universe and can provide precise information about the nature and 
history of the underlying spacetime \cite{martinez3,pierini3,pierini4}. Thus their study may prove useful in constructing the early universe models. 
Motivated by this idea, here we try to explore the possible effects of a string inspired second rank antisymmetric tensor field, 
known as Kalb-Ramond (KR) field, on cosmological scalar particle production as well as on quantum entanglement between the particles produced 
with a hope that the entanglement entropy may provide a possible testbed for the existence of the KR field in our universe 
(see \cite{kalb,callan} for some of the seminal works on Kalb-Ramond field).\\ 

This string-inspired term is motivated by the fact that during the primordial epoch, quantum gravity or string 
theory effects may have a significant imprint on the evolution of the Universe, so in our case we quantify the quantum epoch’s 
imprint on the evolution of the Universe, by using 
this rank two KR antisymmetric tensor field. In general, antisymmetric tensor fields or equivalently p-forms, constitute 
the field content of all superstring models, and in effect these can actually have a realistic impact in the low-energy limit of the theory 
\cite{buchbinder}. 
Apart from the string theory view point, the KR field also plays significant role in many other places as well, some of them are given by :
\begin{itemize}
\item Modified theories of gravity, formulated using twistors, require the inclusion of this antisymmetric tensor field \cite{howe1,howe2}.

\item Attempts to unify gravity and electromagnetism necessitates the inclusion of Kalb-Ramond field in higher-dimensional theories \cite{kubyshin,german}.

\item Spacetime endowed with Kalb-Ramond field becomes optically active exhibiting birefringence \cite{kar1,kar2}. 

\item In \cite{majumdar} an antisymmetric tensor field $B_{\mu\nu}$ identified to be the Kalb-Ramond field was shown to act as the 
source of spacetime torsion. 

\end{itemize}

Most importantly, in the context of cosmology, the KR field energy density is found to decrease as $1/a^6$ (with $a$ being the scale factor of our universe) 
with the expansion of our universe \cite{tp1,tp2,tp3} i.e at a faster rate in comparison to radiation and matter components. 
Thus as the Universe evolves and cools down, the contribution of the KR field on the evolutionary process reduces significantly, 
and at present it almost does not affect the evolution. However the KR field has a significant 
contribution during early universe (when the scale factor is small), 
in particular, it affects the beginning of inflation as well as 
increases the amount of primordial gravitational radiation and hence enlarges the value of tensor to scalar ratio in respect to the case 
when the KR field is absent. The important question that ramains is:

\begin{itemize}
 \item Sitting in present day universe, how do we confirm the existence of the Kalb-Ramond field which has considerably low energy density (with respect 
 to the other components) in our present universe ? 
\end{itemize}

The answer to this question may be encripted in some late time phenomena which carries the information of early universe. One of such phenomena can be 
the ``Cosmological Quantum Entanglement''. Keeping this in mind, here we try to address the possible effects of KR field on cosmological 
particle production as well as on quantum entanglement for a massive scalar field propagating in a four dimensional FRW spacetime.\\
The paper is organized as follows: in section II, we present the model and the evolution of classical fields. Section III is reserved for 
scalar field quantization, calculation of Bogolyubov coefficients, cosmological scalar particle production and quantum entanglement entropy 
between the produced particles in presence of Kalb-Ramond field and their possible consequences. We finally end the paper with some concluding remarks.

\section{The model and the evolution of classical fields}

In the present paper, we are interested on quantum evolution of a massive scalar field in the bckground of FRW spacetime along with a second 
rank antisymmetric tensor field, generally known as Kalb-Ramond (KR) field. In particular, our main goal is to determine how the presence of KR field 
affects the scalar field particle production and the quantum entanglement entropy between the scalar particles. The scalar field 
coupled with KR field action is given by,

\begin{eqnarray}
 S = \int d^4x \sqrt{-g} \bigg[\frac{1}{2}g^{\mu\nu}\partial_{\mu}\partial_{\nu}\Phi - \frac{1}{2}m^2\Phi^2 - \frac{1}{2}\xi R\Phi^2 
 - \frac{1}{12}H_{\mu\nu\alpha}H^{\mu\nu\alpha} - \frac{\alpha}{2}f(\Phi)H_{\mu\nu\alpha}H^{\mu\nu\alpha}\bigg]
 \label{action1}
\end{eqnarray}
where $\Phi$ is the scalar field with $m$ is its mass and $H_{\mu\nu\alpha} (= \partial_{[\mu}B_{\nu\alpha]})$ is the field strength tensor 
of the KR field $B_{\mu\nu}$. Needless to say, first and fourth terms of the above action represent the kinetic terms of th scalar field 
and the KR field respectively. Moreover the scalar field is non-minimally coupled with gravity and also coupled to 
$H_{\mu\nu\alpha}$ via the function $f(\Phi)$. In the absence of KR field, the case, $m = 0$ and $\xi = 1/6$ yields a conformally invariant theory. 
However it is clear that the presence of KR field breaks the conformal symmetry even for $m = 0$ and $\xi = 1/6$ 
because of the coupling function $f(\Phi)$, which has some interesting consequences on particle production and quantum entanglement entropy 
of the scalar field, as will be discussed later.\\
Here the spacetime and the KR field are considered as classical fields while the scalar field ($\Phi(x^{\mu})$) is quantized in this background. 
In this section, we determine the classical evolution of the KR field while the quantization of $\Phi$ 
( coupled with the KR field ) is reserved for the next section.\\
As mentioned earlier, the background classical spacetime is the spatially flat FRW one i.e the metric is given by,

\begin{eqnarray}
 ds^2 = dt^2 - a^2(t)\big[dx^2 + dy^2 + dz^2\big], 
 \label{comoving metric}
\end{eqnarray}
with $t$ and $a(t)$ being the cosmic time and the scale factor respectively. Transforming cosmic time to conformal time ($\eta$) 
by $d\eta = \frac{dt}{a(t)}$, the above spacetime metric can be written as,

\begin{eqnarray}
 ds^2 = a^2(\eta) \big[d\eta^2 - dx^2 - dy^2 - dz^2\big].
 \label{conformal metric}
\end{eqnarray}

It is evident that the FRW spacetime with ($\eta$, x, y, z) coordinate system (known as conformal coordinate ) 
is conformally connected to Minkowski (or flat) spacetime represented by 
the same coordinate system. \\

Before presenting the field equations, we want to note that due to the totally antisymmetric nature, $H_{\mu\nu\alpha}$ 
has four independent components in a four dimensional spacetime, which can be expressed as follows:

\begin{eqnarray}
 H_{012} = h_1~~~~~~~~~~~~,~~~~~~~~~~~~H^{012} = h^1\nonumber\\
 H_{013} = h_2~~~~~~~~~~~~,~~~~~~~~~~~~H^{013} = h^2\nonumber\\
 H_{023} = h_3~~~~~~~~~~~~,~~~~~~~~~~~~H^{023} = h^3\nonumber\\
 H_{123} = h_4~~~~~~~~~~~~,~~~~~~~~~~~~H^{123} = h^4
\label{independent}
\end{eqnarray}
At this stage, it is worth mentioning that due to the presence of the four independent components, the KR field tensor $H_{\mu\nu\alpha}$ can be 
equivalently expressed by a vector field (which has also four independent components in four dimensions) as 
$H_{\mu\nu\alpha} = \varepsilon_{\mu\nu\alpha\beta}f^{-1}(\Phi)\Upsilon^{\beta}$ with $\Upsilon^{\beta}$ being the vector field. 
By virtue of eqn.(\ref{independent}), the off-diagonal Einstein's equations become,

 \begin{eqnarray}
 h_4h^3 = h_4h^2 = h_4h^1 = h_2h^3 = h_1h^3 = h_1h^2 = 0,
 \nonumber
\end{eqnarray}
which have the following solution
\begin{eqnarray}
 h_1 = h_2 = h_3 = 0~~~~~~~,~~~~~~~h_4 \neq 0
 \nonumber
\end{eqnarray}
Thus out of the four independent components of $H_{\mu\nu\alpha}$, only one component i.e $H_{123} = h_4$ comes with non trivial solution (other three 
become trivially zero, which is also in agreement with the isotropic condition of the spacetime). 
This solution along with the equivalence of $H_{\mu\nu\alpha}$ 
with the vector field $\Upsilon^{\beta}$, one finds that $\Upsilon^{\beta}$ has also only one non zero component. 
As a consequence, in a spatially flat FRW spacetime (where the off-diagonal components of Einstein tensor vanish ), 
$\Upsilon^{\beta}$ can be expressed as a derivative of a massless scalar 
field $Z(x^{\mu})$ (i.e $\Upsilon^{\beta} = \partial^{\beta}Z$ where $Z(x)$ is known as axion field), which further relates the KR
field tensor with the axion field in the following way, 

\begin{eqnarray}
 H^{\mu\nu\lambda} = \epsilon^{\mu\nu\lambda\beta} f^{-1}(\Phi) \partial_{\beta}Z.
 \label{KR to axion}
\end{eqnarray}

With the help of eqn.(\ref{KR to axion}), the equation of motion for KR field can be obtained as,

\begin{eqnarray}
 \frac{d}{d\eta}\bigg[f^{-1}(\Phi)\frac{dZ}{d\eta}\bigg] + 2\frac{\dot{a}}{a}\bigg[f^{-1}(\Phi)\frac{dZ}{d\eta}\bigg] = 0.
 \label{axion equation}
\end{eqnarray}
The above differential equation can be integrated once to yeild,

\begin{eqnarray}
 \frac{dZ}{d\eta} \propto f(\Phi)/a^2.
 \label{axion solution}
\end{eqnarray}
Using the above expression, we determine the KR field energy density ($\rho_{KR}$) in terms of the scale factor $a(\eta)$ as follows,

\begin{eqnarray}
 \rho_{KR} = \frac{1}{2}h_4h^4 = \frac{1}{2}g^{00} \bigg[f^{-1}(\Phi)\frac{dZ}{d\eta}\bigg]^2 = \frac{h_0}{a^6}
 \label{KR energy density}
\end{eqnarray}
with $h_0$ being an integration constant which must take only positive values in order to get 
a real valued solution for $h_4(t)$. In addition, eqn.(\ref{KR energy density}) clearly indicates that the energy 
density of the KR field ($\rho_{KR}$) is proportional to $1/a^6$ and in effect, $\rho_{KR}$ decreases as the 
Universe expands, in a faster rate in comparison to matter ($\propto 1/a^3$) and radiation ($\propto 1/a^4$) energy densities respectively.\\
With the classical evolution of KR field in hand, we now discuss the possible effects of KR field on the scalar field quantization, 
scalar particle production, entanglement entropy between the scalar particles etc. \\

\section{Scalar field quantization, Bogolyubov coefficients, scalar particle production and entanglement entropy in the presence of KR field}

The action in eqn.(\ref{action1}) leads to the scalar field ($\Phi$) equation as,
\begin{eqnarray}
 \bigg[\Box_{g} + m^2 + \xi R\bigg]\Phi + \alpha \rho_{KR} f'(\Phi) = 0,
 \label{scalar field equation1}
\end{eqnarray}

where $f'(\Phi) = \frac{df}{d\Phi}$ and recall $\rho_{KR} = \frac{1}{2}H_{\mu\nu\lambda}H^{\mu\nu\lambda} = \frac{1}{2}h_4h^4$. 
The coupling $f(\Phi)$ is the term that carries the imprint of the KR field on scalar field quantization. Without the 
coupling term (i.e for $\alpha = 0$), the scalar field quantization is unaffected by the KR field evolution, which is 
also evident from eqn.(\ref{scalar field equation1}). However the scalar field is 
propagating in FRW spacetime charted by conformal coordinate system, for which the box operator takes the form

\begin{eqnarray}
 \Box_{g}&=&\frac{1}{\sqrt{-g}} \partial_{\mu}\bigg[\sqrt{-g}g^{\mu\nu}\partial_{\nu}\bigg]\nonumber\\
 &=&\frac{1}{a^4} \bigg[\frac{\partial}{\partial \eta}\bigg(a^4g^{00}\frac{\partial}{\partial \eta}\bigg) 
 + \frac{\partial}{\partial x}\bigg(a^4g^{11}\frac{\partial}{\partial x}\bigg) 
 + \frac{\partial}{\partial y}\bigg(a^4g^{22}\frac{\partial}{\partial y}\bigg) 
 + \frac{\partial}{\partial z}\bigg(a^4g^{33}\frac{\partial}{\partial z}\bigg)\bigg]\nonumber\\
 &=&\frac{1}{a^2} \bigg[\frac{\partial^2}{\partial\eta^2} - \nabla^2 + \frac{2\dot{a}}{a}\frac{\partial}{\partial\eta}\bigg].
 \label{dot_conformal}
\end{eqnarray}
where an overdot represents derivative with respect to the conformal time i.e $\frac{d}{d\eta}$. Expanding the field $\Phi$ in Fourier modes as,

\begin{eqnarray}
 \Phi(\vec{x},\eta)&=&\int d^3k \Phi_{\vec{k}}(\vec{x},\eta)\nonumber\\
 &=&\int d^3k \bigg[\frac{1}{(2\pi)^{3/2}} e^{i\vec{k}.\vec{x}} \frac{\chi_k(\eta)}{a(\eta)}\bigg],
 \label{fourier decomposition}
\end{eqnarray}
where we also introduce the auxiliary field $\chi_k(\eta) = a(\eta)\Phi(\eta)$. With such auxiliary field, we simplify the term 
$\Box_{g}\Phi_{\vec{k}}$ to yeild -
\begin{eqnarray}
 \Box_{g}\Phi_{\vec{k}} = e^{i\vec{k}.\vec{x}} \bigg[\frac{1}{a^3}\big(\ddot{\chi}_{k} + k^2\chi_k\big) - \frac{1}{6}R(\eta)\chi_{k}\bigg],
 \label{box simplification}
\end{eqnarray}

with $R(\eta) = 6\ddot{a}/a^3$ being the scalar curvature of the spacetime and $\ddot{\chi}_{k} = \frac{d^2\chi_{k}}{d\eta^2}$ 
(recall the overdot stands for $\frac{d}{d\eta}$ as mentioned after Eq.(\ref{dot_conformal})). 
Now in order to get an explicit form of the scalar field equation, the 
coupling function $f(\Phi)$ is taken as quadratic one i.e
\begin{eqnarray}
 f(\Phi) = \Phi^2/2.
 \label{form of f}
\end{eqnarray}
Such quadratic form of the coupling function leads to the scalar field equation as similar to harmonic oscillator like equation (with time dependent 
frequency, as we will see below in eqn.(\ref{scalar field equation2})), which in turn makes the scalar field quantization easier. 
Plugging the expression of $\Box_{g}\Phi_{\vec{k}}$ into eqn.(\ref{scalar field equation1}) and using the quadratic form of $f(\Phi)$, one obtains 
the following equations of motion for the modes $\chi_k(\eta)$ :

\begin{eqnarray}
 \ddot{\chi}_k + \bigg[k^2 + \big(\xi - \frac{1}{6}\big) R(\eta)a^2(\eta) + m^2a^2(\eta) + \frac{\alpha h_0}{a^4}\bigg]\chi_k(\eta) = 0,
 \label{scalar field equation2}
 \end{eqnarray}
where we use the solution of $\rho_{KR} = h_0/a^6$ (see eqn.(\ref{KR energy density})).\\ 
It is evident that the mode functions get decoupled 
in the above equation of motion. Such decoupling of the field modes hinges on the separation 
of the time coordinate in the Klein-Gordon equation and on the Fourier expansion of the field $\Phi$ 
through the eigenfunctions of the spatial Laplace operator at a fixed time. It may be mentioned that 
the field equations in a general spacetime are not separable. 
In such cases, the mode decoupling cannot be performed explicitly and quantization is difficult. However this 
is not the subject of our present paper and thus 
without going into details of the quantization in a general spacetime, we keep our focus on that in the expanding FRW spacetime in presence of KR field.

\subsection{Quantization of the scalar field}

In order to quantize the scalar field, $\chi_k(\eta)$ is replaced by the field operator i.e $\hat{\chi}_k(\eta)$ and 
$\hat{\Psi}_k(\vec{x},\eta) = e^{i\vec{k}.\vec{x}}\hat{\chi}_k(\eta)$ satisfies the following equal time commutation relation,

\begin{eqnarray}
 \bigg[\hat{\Psi}_k(\vec{x},\eta) , \hat{\pi}_k(\vec{x},\eta)\bigg] = i\delta(\vec{x} - \vec{y}),
 \label{commutation}
\end{eqnarray}

with $\hat{\pi}_k(\vec{x},\eta) = \frac{d\hat{\Psi}_k(\vec{x},\eta)}{d\eta}$ be the canonical momentum conjugate to 
$\hat{\Psi}_k(\vec{x},\eta)$. The quantum Hamiltonian comes with the following expression,
\begin{eqnarray}
 \hat{H}_k(\eta) = \frac{1}{2}\int d^3x \bigg[\hat{\pi}_k^2 + (\nabla\hat{\Psi}_k)^2 + m_{eff}^2(\eta) \hat{\Psi}_k^2(\vec{x},\eta)\bigg]
 \label{hamiltonian}
\end{eqnarray}
with $m_{eff}^2(\eta) = \big(\xi - \frac{1}{6}\big) R(\eta)a^2(\eta) + m^2a^2(\eta) + \frac{\alpha h_0}{a^4}$, known as effective mass 
of the field mode. The Hamiltonian $\hat{H}_k(\eta)$ resembles with a harmonic oscillator Hamiltonian except the fact 
that the mass term here explicitly depends on time, which in turn makes the Hamiltonian explicit time dependent. This time dependence 
arises due to the interaction of the scalar field with the background time dependent gravitational field (i.e FRW spacetime). The expression of 
$\hat{H}$ along with the Fourier decomposition of $\hat{\Psi}_k(\vec{x},\eta)$ lead to the Heisenberg equation for $\hat{\chi}_k(\eta)$ as follows,
\begin{eqnarray}
 \frac{d^2\hat{\chi}_k}{d\eta^2} 
 + \bigg[k^2 + \big(\xi - \frac{1}{6}\big) R(\eta)a^2(\eta) + m^2a^2(\eta) + \frac{\alpha h_0}{a^4}\bigg]\hat{\chi}_k(\eta) = 0 ,
 \label{scalar field equation3}
\end{eqnarray}

where we use the explicit form of $m_{eff}^2(\eta)$. It is evident that the form of eqn.(\ref{scalar field equation3}) resembles with that 
of eqn.(\ref{scalar field equation2}). However this is expected as Heisenberg equation resembles to the corresponding classical field equation with 
the replacement of the ``classical field'' by the ``field operator''.\\
Till now, we did not consider any particular form of the scale factor ($a(\eta)$), but in order 
to proceed further, we do need a certain form of $a(\eta)$ and thus we choose a suitable form of the background FRW spacetime as follows,
\begin{eqnarray}
 a(\eta) = 1 - \frac{\sigma^2}{2\big(\eta^2 + \sigma^2\big)},
 \label{scale factor}
\end{eqnarray}

with $\sigma$ being the model parameter. The above form of the scale factor corresponds to a non-singular symmetric bounce at $\eta = 0$. 
Such form of the scale factor has also been used earlier in \cite{pierini3} 
to study the possible effects of spacetime anisotropy on cosmological 
entanglement. At this stage it deserves mention that here, in the present paper, 
we choose this certain form of the scale factor in order to make the calculations easier. 
However we will show that our final argument regarding the testbed of KR field through 
entanglement entropy remains valid for the class of scale factor (irrespective of a particular form) which exhibits a symmetric bounce.\\
Eqn.(\ref{scale factor}) leads to $a(\pm \infty) = 1$ which in turn admits an asymptotically flat spacetime. In order 
to reveal the asymptotic nature of spacetime more clearly, we rewrite eqn.(\ref{scalar field equation3}) in the following form,

\begin{eqnarray}
 \frac{d^2\hat{\chi}_k}{d\eta^2} + \bigg[\omega_k^2 - V_k(\eta)\bigg]\hat{\chi}_k(\eta) = 0,
 \label{scalar field equation4}
\end{eqnarray}
where $\omega_k^2 = k^2 + m^2 + \alpha h_0$ and
\begin{eqnarray}
 V_k(\eta) = m^2\big[1 - a^2(\eta)\big] - \big(\xi - \frac{1}{6}\big)R(\eta)a^2(\eta) + \alpha h_0\bigg[1 - \frac{1}{a^4(\eta)}\bigg].
 \label{potential}
\end{eqnarray}

The above expression clearly demonstrates that for a flat background spacetime (i.e for $a(\eta) = 1$), 
$V_k(\eta)$ becomes trivially zero. In the present context, 
$V_k(\eta)$ attains a non-zero value because the FRW spacetime is a ``curved'' spacetime, and moreover it encodes the information of the 
interaction of the scalar field with the background time dependent gravitational field. This is the reason that $V_k(\eta)$ is generally known as 
potential term or interaction term. Eqn.(\ref{potential}) further entails that $V_k(\eta)$ consists of three parts - 
the first one is proportional to $m^2$, the second one is proportional 
to $(\xi-1/6)$ while the third part arises due to the presence of KR field and depends on $h_0$. 
The term $R(\eta)a^2(\eta) \sim \frac{3\eta^2 - \sigma^2}{(\eta^2 + \sigma^2)^2(2\eta^2 + \sigma^2)}$ goes to zero for $\eta \rightarrow \pm\infty$ 
and thus the interaction term $V_k(\eta)$ also approaches to zero asymptotically. Thereby eqn.(\ref{scalar field equation4}) 
clearly demonstrates that the mode operator $\hat{\chi}_k(\eta)$ 
obeys the flat spaceime like equation in the regime $\eta \rightarrow \pm\infty$ with $\omega_k^2 = k^2 + m^2 + \alpha h_0$. 
As a consequence - if $\zeta_k^{(in)}(\eta)$ and $\zeta_k^{(out)}(\eta)$ are the mode solutions of eqn.(\ref{scalar field equation4}), 
which are consistent with the correct vacuum of the scalar field at $\eta \rightarrow \mp\infty$ respectively, 
then such mode solutions come with the following asymptotic behaviour,
\begin{eqnarray}
 \zeta_k^{(in)}(\eta \rightarrow -\infty) = \frac{1}{\sqrt{2\omega_k}}e^{-i\omega_k\eta}\nonumber\\
 \zeta_k^{(out)}(\eta \rightarrow +\infty) = \frac{1}{\sqrt{2\omega_k}}e^{-i\omega_k\eta}
 \label{correct mode function}
\end{eqnarray}

$\zeta_k^{(in)}(\eta)$ and $\zeta_k^{(out)}(\eta)$ are known as ``in-mode'' and ``out-mode'' solution respectively. 
Moreover $\hat{\chi}_k(\eta)$ can be written as linear combination (as eqn.(\ref{scalar field equation4}) is linear in $\chi_k$) 
of these mode solutions as follows : 
\begin{eqnarray}
 \hat{\chi}_k(\eta)&=&\hat{a}_{\vec{k}} \zeta_k^{(in)*}(\eta) + \hat{a}_{-\vec{k}}^{+} \zeta_k^{(in)}(\eta)\nonumber\\
 &=&\hat{b}_{\vec{k}} \zeta_k^{(out)*}(\eta) + \hat{b}_{-\vec{k}}^{+} \zeta_k^{(out)}(\eta),
 \label{mode decomposition}
\end{eqnarray}

where $\hat{a}_{\vec{k}}$ and $\hat{b}_{\vec{k}}$ are operator coefficients which obey commutation relations like - 
$[\hat{a}_{\vec{k}} , \hat{a}^+_{\vec{k'}}] = \delta(\vec{k} - \vec{k'})$ , $[\hat{b}_{\vec{k}} , \hat{b}^+_{\vec{k'}}] = \delta(\vec{k} - \vec{k'})$ 
and all other commutators are zero. Such operator coefficients actually represent the annihilation operator for in ($\eta \rightarrow -\infty$) 
and out ($\eta \rightarrow +\infty$) region respectively. Thus the ``in-vacuum`` ($| 0_{in} \textgreater$) and ''out-vacuum`` 
($| 0_{out} \textgreater$) state of the scalar field are defined as,  
\begin{eqnarray}
 | 0_{in} \textgreater~~~~:~~~~~~~~~~~~\hat{a}_{\vec{k}}| 0_{in} \textgreater = 0 ~~~\forall \vec{k}, \nonumber\\
 | 0_{out} \textgreater~~~~:~~~~~~~~~~~~\hat{b}_{\vec{k}}| 0_{out} \textgreater = 0 ~~~\forall \vec{k}.
 \label{vacuum}
\end{eqnarray}

It may be mentioned that $| 0_{in} \textgreater$ and $| 0_{out} \textgreater$ are two different states in the Fock space. The interaction of 
the scalar field with the background FRW spacetime makes the scalar field Hamiltonian explicitly time dependent (see eqn.(\ref{hamiltonian})) 
which in turn causes the difference in the asymptotic vacuum states of the scalar field in the Fock space.\\ 
The prefactor $\frac{1}{\sqrt{2\omega_k}}$ in the asymptotic behaviour of ''in-mode`` and ''out-mode`` solutions (see eqn.(\ref{correct mode function})) 
ensures their normalization conditions as,
\begin{eqnarray}
 \bigg(\zeta_k^{(in)}(\eta) , \zeta_k^{(in)}(\eta)\bigg) = 1 , \nonumber\\
 \bigg(\zeta_k^{(in)*}(\eta) , \zeta_k^{(in)*}(\eta)\bigg) = -1 , \nonumber\\
 \bigg(\zeta_k^{(in)}(\eta) , \zeta_k^{(in)*}(\eta)\bigg) = 0.
 \label{normalization condition}
\end{eqnarray}

and the same for the out modes also, where $\bigg(\zeta_1 , \zeta_2\bigg)$ is the inner product of the corresponding functions given by
$\bigg(\zeta_1 , \zeta_2\bigg) = \zeta_1 \partial_{\eta}\zeta_2^{*} - \zeta_2^* \partial_{\eta}\zeta_1$. Using eqn.(\ref{correct mode function}), 
the integral form of the differential 
equation (\ref{scalar field equation4}) can be expressed as,

\begin{eqnarray}
 \zeta_k^{(in)}(\eta) = \frac{1}{\sqrt{2\omega_k}}e^{-i\omega_k\eta} 
 + \frac{1}{i\omega_k}\int_{-\infty}^{\eta} d\eta_1 V_k(\eta_1)\bigg[e^{i\omega(\eta-\eta_1)} - e^{-i\omega(\eta-\eta_1)}\bigg]\chi_k(\eta_1)
 \label{in mode solution}
\end{eqnarray}
Eqn.(\ref{in mode solution}) reveals that 
$\zeta_k^{(in)}(\eta)$ starts with the plane wave solution at past infinity (also shown in eqn.(\ref{correct mode function})), but 
the presence of the interaction term $V_k(\eta)$ causes the deviation of the in-mode solution from the plane wave form with the expansion 
of our universe. This in turn changes the annihilation operator during cosmic evolution, which is reflected through the fact that 
the vacuum states of the scalar field in the two asymptotic regimes ($\eta \rightarrow \pm\infty$) are different.

\subsection{Bogolyubov coefficients}
As $\big[\zeta_k^{(in)}(\eta)$ , $\zeta_k^{(in)*}(\eta)\big]$ and $\big[\zeta_k^{(out)}(\eta)$ , $\zeta_k^{(out)*}(\eta)\big]$ 
are the two basis sets in the field space, 
one can express $\zeta_k^{(in)}(\eta)$ as the linear combination of the other basis sets as, 
\begin{eqnarray}
 \zeta_k^{(in)}(\eta) = \alpha_k \zeta_k^{(out)}(\eta) + \beta_k \zeta_k^{(out)*}(\eta)~~\forall \eta,
 \label{bogolyubov 1}
\end{eqnarray}

with $\alpha_k$ and $\beta_k$ are known as the Bogolyubov coefficients. Since 
the above expression is valid for entire range of $\eta$, we can immediately write 
\begin{eqnarray}
\zeta_k^{(in)}(\eta \rightarrow \infty) = \alpha_k \zeta_k^{(out)}(\eta \rightarrow \infty) + \beta_k \zeta_k^{(out)*}(\eta \rightarrow \infty).
\label{bogolyubov 2}
\end{eqnarray}

Imposing the normalization conditions on eqn.(\ref{bogolyubov 2}), the Bogolyubov coefficients can be obtained as the inner 
products of $in$ and $out$ modes as

\begin{eqnarray}
 \alpha_k&=&\bigg(\zeta_k^{(in)}(\eta \rightarrow \infty) , \zeta_k^{(out)}(\eta \rightarrow \infty)\bigg), \nonumber\\
 \beta_k&=&- \bigg(\zeta_k^{(in)}(\eta \rightarrow \infty) , \zeta_k^{(out)*}(\eta \rightarrow \infty)\bigg).
 \label{bogolyubov 3}
\end{eqnarray}
Using the integral form of $\zeta_k^{(in)}(\eta)$ (see eqn.(\ref{in mode solution})) along with the condition $V_k(\eta \rightarrow \pm\infty)$, the 
above expression can be simplified to yeild

\begin{eqnarray}
 \alpha_k&=&1 + i\int_{-\infty}^{\infty} d\eta \zeta_k^{(out)*}(\eta \rightarrow \infty) V_k(\eta) \zeta_k^{(in)}(\eta).\nonumber\\
 \beta_k&=&-i\int_{-\infty}^{\infty} d\eta \zeta_k^{(out)}(\eta \rightarrow \infty) V_k(\eta) \zeta_k^{(in)}(\eta).
 \label{bogolyubov 4}
\end{eqnarray}

Clearly, in the absence of the interaction term (i.e for $V-k(\eta) = 0$ which occurs in flat spacetime, as mentioned earlier), 
the Bogolyubov coefficients have the values $\alpha_k = 1$ and $\beta_k = 0$ respectively. To solve eqn.(\ref{bogolyubov 4}), we resort to an iterative 
procedure. The lowest order gives,
\begin{eqnarray}
 \zeta_k^{(in)}(\eta) = \frac{1}{\sqrt{2\omega_k}}e^{-i\omega_k\eta}
 \nonumber
\end{eqnarray}

and consequently the Bogolyubov coefficients have the following expressions :
\begin{eqnarray}
 \alpha_k&=&1 + \frac{i}{2\omega_k} \int_{-\infty}^{\infty} d\eta V_k(\eta)\nonumber\\
 &=&1 + \frac{i}{2\omega_k} \int_{-\infty}^{\infty} d\eta \bigg[m^2\bigg(a^2(\infty) - a^2(\eta)\bigg) 
 - \big(\xi - \frac{1}{6}\big)R(\eta)a^2(\eta) + \alpha h_0\bigg(\frac{1}{a^4(\infty)} 
 - \frac{1}{a^4(\eta)}\bigg)\bigg]\nonumber\\
 &=&1 + \alpha_k^{(mass)} + \alpha_k^{(coup)} + \alpha_k^{(KR)}
 \label{bogolyubov 5a}
\end{eqnarray}

and 
\begin{eqnarray}
 \beta_k&=&-\frac{i}{2\omega_k} \int_{-\infty}^{\infty} d\eta e^{-2i\omega\eta} V_k(\eta)\nonumber\\
 &=&\beta_k^{(mass)} + \beta_k^{(coup)} + \beta_k^{(KR)},
 \label{bogolyubov 5b}
\end{eqnarray}
where $\alpha_k^{(mass)}$ ($\beta_k^{(mass)}$), $\alpha_k^{(coup)}$ ($\beta_k^{(coup)}$) and $\alpha_k^{(KR)}$ ($\beta_k^{(KR)}$) 
are proportional to $m^2$ (scalar field mass), $(\xi-1/6)$ (curvature coupling) and $h_0$ (KR field energy density) respectively. Thereby 
the Bogolyubov coefficients split into three parts, just like $V_k(\eta)$ in eqn.(\ref{potential}). Using the form of $a(\eta)$ 
(see eqn.(\ref{scale factor})), we integrate the above expressions to determine the explicit expressions for various 
parts of $\alpha_k$ and $\beta_k$ as follows,

\begin{eqnarray}
 \alpha_k^{(mass)}&=&\frac{i}{2\omega_k} \int_{-\infty}^{\infty} d\eta~m^2\big[1 - a^2(\eta)\big]\nonumber\\
 &=&i \frac{7m^2\sigma\pi}{16\sqrt{k^2 + m^2 + \alpha h_0}},
 \label{bogolyubov 6a}
\end{eqnarray}

\begin{eqnarray}
 \alpha_k^{(coup)}&=&-\frac{i}{2\omega_k} \big(\xi - \frac{1}{6}\big) \int_{-\infty}^{\infty} d\eta~R(\eta)a^2(\eta)\nonumber\\
 &=&i\frac{6(7 - 5\sqrt{2}) \big(\xi - \frac{1}{6}) \pi}{\sigma \sqrt{k^2 + m^2 + \alpha h_0}},
 \label{bogolyubov 6b}
\end{eqnarray}

\begin{eqnarray}
 \alpha_k^{(KR)}&=&\frac{i}{2\omega_k} \alpha h_0 \int_{-\infty}^{\infty} d\eta \bigg[1 - \frac{1}{a^4(\eta)}\bigg]\nonumber\\
 &=&-i \frac{141\alpha h_0\sigma \pi}{32\sqrt{2}\sqrt{k^2 + m^2 + \alpha h_0}},
\label{bogolyubov 6c}
\end{eqnarray}

and

\begin{eqnarray}
 \beta_k^{(mass)}&=&-\frac{i}{2\omega_k} \int_{-\infty}^{\infty} d\eta~e^{-2i\omega\eta} m^2\big[1 - a^2(\eta)\big]\nonumber\\
 &=&i\frac{m^2\sigma\pi (2\sigma\sqrt{k^2+m^2+\alpha h_0} - 7)e^{-2\omega\sigma}}{16\sqrt{k^2+m^2+\alpha h_0}},
 \label{bogolyubov 7a}
\end{eqnarray}

\begin{eqnarray}
 \beta_k^{(coup)}&=&\frac{i}{2\omega_k} \big(\xi - \frac{1}{6}\big) \int_{-\infty}^{\infty} d\eta~e^{-2i\omega\eta} R(\eta)a^2(\eta)\nonumber\\
 &=&i\frac{\big(\xi - \frac{1}{6}\big)\pi}{16\sigma\sqrt{k^2 + m^2 + \alpha h_0}} 
 \bigg[(384\sigma\omega + 672)e^{-2\omega\sigma} - 480\sqrt{2}e^{-\sqrt{2}\omega\sigma}\bigg],
 \label{bogolyubov 7b}
\end{eqnarray}

\begin{eqnarray}
 \beta_k^{(KR)}&=&-\frac{i}{2\omega_k} \alpha h_0 \int_{-\infty}^{\infty} d\eta~e^{-2i\omega\eta}\bigg[1 - \frac{1}{a^4(\eta)}\bigg]\nonumber\\
 &=&i\frac{\alpha h_0\sigma \pi e^{-\sqrt{2}\omega\sigma}}{192\sqrt{k^2 + m^2 + \alpha h_0}} 
 \bigg[423\sqrt{2} + 462\sigma\omega + 60\sqrt{2}\sigma^2\omega^2 + 4\sigma^3\omega^3\bigg],
 \label{bogolyubov 7c}
\end{eqnarray}

where $\omega = \sqrt{k^2 + m^2 + \alpha h_0}$. The quantities that we actually 
need for the purpose of scalar particle production, entanglement entropy are $\big|\alpha_k\big|^2$ and $\big|\beta_k\big|^2$. 
These quantities can be determined as
\begin{eqnarray}
 \big|\alpha_k\big|^2 = 1 + \frac{\pi^2}{(k^2 + m^2 + \alpha h_0)} \bigg[\frac{7m^2\sigma}{16} 
 + \frac{6(7 - 5\sqrt{2}) \big(\xi - \frac{1}{6})}{\sigma} - \frac{141\alpha h_0\sigma}{32\sqrt{2}}\bigg]^2
 \label{bogolyubov 6d}
\end{eqnarray}
and

\begin{eqnarray}
 \big|\beta_k\big|^2 = \frac{\pi^2}{(k^2 + m^2 + \alpha h_0)} 
 &\bigg[&\frac{m^2\sigma (2\sigma\sqrt{k^2+m^2+\alpha h_0} - 7)e^{-2\omega\sigma}}{16} 
 + \frac{\big(\xi - \frac{1}{6}\big)}{16\sigma} 
 \bigg((384\sigma\omega + 672)e^{-2\omega\sigma} - 480\sqrt{2}e^{-\sqrt{2}\omega\sigma}\bigg)\nonumber\\ 
 &+&\frac{\alpha h_0\sigma e^{-\sqrt{2}\omega\sigma}}{192} 
 \bigg(423\sqrt{2} + 462\sigma\omega + 60\sqrt{2}\sigma^2\omega^2 + 4\sigma^3\omega^3\bigg)\bigg]^2 .
 \label{bogolyubov 7d}
\end{eqnarray}

It is clear that the conditions $m= 0$ (massless scalar field) and $\xi = 1/6$ (conformal coupling) do not yield $\big|\alpha_k\big|^2 = 1$ and 
$\big|\beta_k\big|^2 = 0$ : unlike the case when the KR field is absent. This is the consequence of the fact that the presence of the KR field 
(actually the coupling between the scalar field and the KR field) spoils the conformal symmetry of a massless scalar field propagating in FRW spacetime. 
It has interesting effects on scalar particle production as well as on entanglement entropy, as will be discussed in the next sections.

\subsection{Scalar particle production}
The interaction of the scalar field inflicts an energy exchange between the background classical fields ( i.e from gravitational 
field and the KR field ) and the scalar field, which manifests as the scalar particle production. The energy required to excite a 
scalar particle (from vacuum) having momentum 
$\vec{k}$ is given by $\omega_k = \sqrt{k^2 + m^2 + \alpha h_0}$ which clearly indicates that a larger amount of energy is required to create a scalar 
particle in comparison with the case when the KR field is absent. In this section, we calculate the particle number density of the scalar field 
starting from ''infinite-past-vacuum-state`` i.e from $| 0_{in} \textgreater$ and explore the possible effects of KR field. The state of 
the scalar field in the in-region can be written as,
\begin{eqnarray}
 | state~,~-\infty~\textgreater = | 0_{in}\textgreater .
 \label{production1a}
\end{eqnarray}

\begin{figure}
\begin{center}
 \centering
 \includegraphics[width=3.0in,height=2.0in]{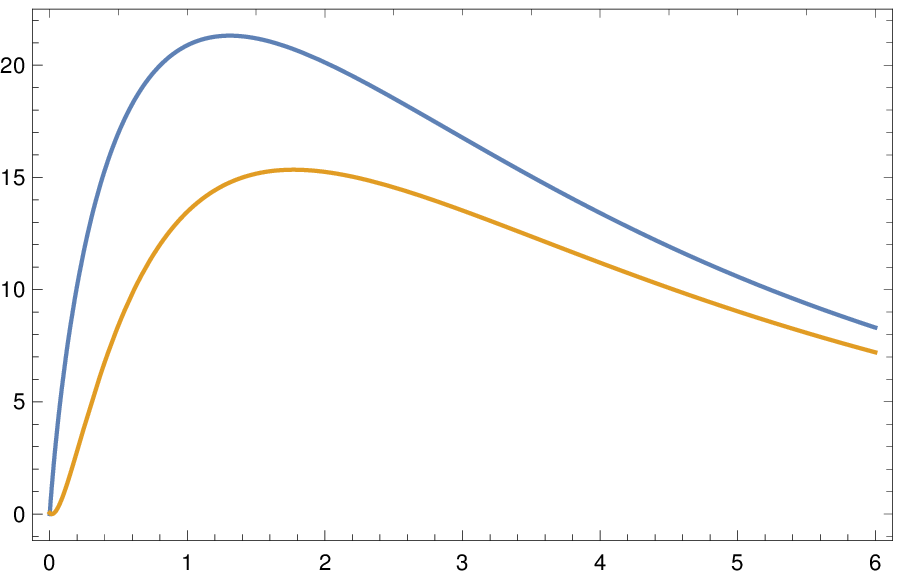}
 \includegraphics[width=3.0in,height=2.0in]{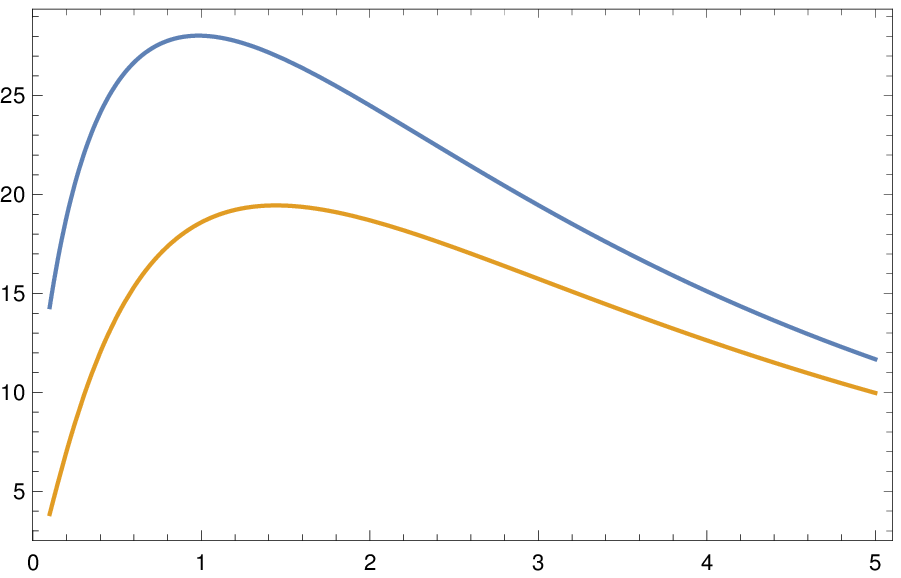}
 \caption{$<n_k>_{out}$ (along y axis) vs. $h_0$ (KR field energy density, along x axis). 
 Left figure : for $\xi = 1/6$ ; $k = 0.01$, $m = 0$ (upper figure) and $k = 0.01$, 
 $m = 0.5$ (lower figure). Right figure : for $\xi = 0$ ; $k = 0.01$, $m = 0$ (upper figure) and $k = 0.01$, $m = 0.5$ (lower figure).}
 \label{plot1}
\end{center}
\end{figure}

Since we are working in the Heisenberg picture, the field state at future infinity is given by $| state~,~+\infty~\textgreater = | 0_{in}\textgreater$. 
Recall, $a_{\vec{k}}$ and $b_{\vec{k}}$ are the annihilation operators for the ''in`` and ''out`` regions respectively. With this information, we 
calculate the particle number density (with momentum $\vec{k}$) in the two asymptotic regimes as,
\begin{eqnarray}
 <n_k>_{in}&=&\textless~state~,~-\infty|~a_{\vec{k}}^+a_{\vec{k}}~|~state~,~-\infty\textgreater 
 = \textless~0_{in}|~a_{\vec{k}}^+a_{\vec{k}}~|~0_{in}\textgreater\nonumber\\
 &=&0
 \label{production2a}
\end{eqnarray}
and
\begin{eqnarray}
 <n_k>_{out}&=&\textless~state~,~+\infty|~b_{\vec{k}}^+b_{\vec{k}}~|~state~,~+\infty\textgreater 
 = \textless~0_{in}|~b_{\vec{k}}^+b_{\vec{k}}~|~0_{in}\textgreater\nonumber\\
 &=&\textless~0_{in}|~\big(\alpha_ka_{\vec{k}}^+ - \beta_ka_{-\vec{k}}\big)\big(\alpha_k^*a_{\vec{k}} - \beta_k^*a_{-\vec{k}}^{+}\big)~|~0_{in}\textgreater 
 = \big|\beta_k\big|^2\nonumber\\
 &=&\frac{\pi^2}{(k^2 + m^2 + \alpha h_0)} 
 \bigg[\frac{m^2\sigma (2\sigma\sqrt{k^2+m^2+\alpha h_0} - 7)e^{-2\omega\sigma}}{16} 
 + \frac{\big(\xi - \frac{1}{6}\big)}{16\sigma} 
 \bigg((384\sigma\omega + 672)e^{-2\omega\sigma} - 480\sqrt{2}e^{-\sqrt{2}\omega\sigma}\bigg)\nonumber\\ 
 &+&\frac{\alpha h_0\sigma e^{-\sqrt{2}\omega\sigma}}{192} 
 \bigg(423\sqrt{2} + 462\sigma\omega + 60\sqrt{2}\sigma^2\omega^2 + 4\sigma^3\omega^3\bigg)\bigg]^2,
 \label{production2b}
\end{eqnarray}

respectively, where we use eqn.(\ref{bogolyubov 7d}). It is clearly evident that the scalar field evolves from a ''zero particle state`` to 
a ''non-zero particle state``. The energy required for such particle production comes from the Kalb-Ramond and the gravitational field. 
Eqn.(\ref{production2b}) reveals that $<n_k>_{out}$ goes as $e^{-|\vec{k}|}/k^2$ for large $|\vec{k}|$, which ensures that the total number of produced 
particles i.e $\int d^3\vec{k} <n_k>_{out}$ yields a finite value.\\
In order to investigate the possible effects of KR field on scalar particle production, we give the plots of $<n_k>_{out}$ vs. $\alpha h_0$ 
for $\xi = 1/6$ (conformal coupling) and $\xi = 0$ (weak coupling), see Figure [\ref{plot1}].

The figures demonstrate the following informations:

\begin{itemize}
 \item In the conformal coupling case : no particle production occurs for a massless scalar field in the absence of Kalb-Ramond 
 field (i.e for $h_0 = 0$), however the presence of KR field (i.e $h_0 \neq 0$) leads to a non-zero value of $<n_k>_{out}$ even for $m = 0$. 
 This is a consequence of the fact 
 that the presence of the KR field breaks the conformal symmetry of the massless scalar field even for $\xi = 1/6$ in four dimensional context, which is 
 clearly evident from action (\ref{action1}). On the other hand, in the weak coupling case, $<n_k>_{out}$ is non-zero for all values of 
 $h_0$ ($h_0 \geq 0$), as expected.
 
 \item Irrespective of conformal or weak coupling, $<n_k>_{out}$ has a maximum at a certain value of $h_0$. However here we show 
 that the class of symmetric bouncing scale factor (irrespective of any specific form) leads to a maxima of $<n_k>_{out}$ 
 at a finite value of $h_0$. In order to explore this, following 
 we determine the rate of change of $<n_k>_{out}$, with respect to $\alpha h_0$, 
 in the limits - $\frac{\alpha h_0}{(k^2 + m^2)} \ll 1$ (i.e for low KR field energy density) and 
 $\frac{\alpha h_0}{(k^2 + m^2)} \gg 1$ (i.e for large KR field energy density) respectively. Using eqn.(\ref{bogolyubov 5b}), we obtain the explicit 
 expression of $|\beta_k\big|^2$ for a symmetric universe, as follows:
 
 \begin{eqnarray}
  |\beta_k\big|^2 = \frac{1}{4\omega_k^2} \bigg[I_1 - I_2 + I_3\bigg]^2
  \label{new1} 
  \end{eqnarray}
  
  where the quantities within the square paranthesis have the following expressions:
  \begin{eqnarray}
   I_1&=&2m^2 \int_0^{\infty}d\eta\cos{(2\omega\eta)} \big[a^2(\infty) - a^2(\eta)\big]\nonumber\\
   I_2&=&2(\xi - 1/6) \int_0^{\infty}d\eta\cos{(2\omega\eta)} R(\eta)a^2(\eta)\nonumber\\
   I_3&=&2\alpha h_0 \int_0^{\infty}d\eta\cos{(2\omega\eta)} \bigg[\frac{1}{a^4(\infty)} - \frac{1}{a^4(\eta)}\bigg]
   \label{new2}
  \end{eqnarray}
  
  Due to the condition $a(\eta)=a(-\eta)$ (for which the curvature $R(\eta)=6\ddot{a}/a^3$ also becomes symmetric), 
  the integration limit in $I_1$, $I_2$ and $I_3$ becomes zero to infinity and the other integrals involving 
  $\sin{(2\omega\eta)}$ vanish. Now in the expanding regime 
  i.e for $0 \leq \eta < \infty$, the quantity $\bigg[\frac{1}{a^4(\infty)} - \frac{1}{a^4(\eta)}\bigg]$ is negative and goes to zero asymptotically. 
  Thereby the magnitude $\bigg|\frac{1}{a^4(\infty)} - \frac{1}{a^4(\eta)}\bigg|$ acquires the maximum value at $\eta = 0$ and decreases monotonically 
  with the expansion of our universe. On the other hand, $\cos{(2\omega\eta)}$ starts with a positive value from $\eta = 0$. Thus the negative contribution 
  of the integrand in $I_3$ exceeds than that of the positive contribution. As a consequence the integral $I_3$ must be a negative quantity. Similar 
  argument leads to $I_1 > 0$ and $I_2 \leq 0$ (the equality sign is for the conformal coupling). These 
  informations will be useful later. With the help of eqn.(\ref{new1}), we determine the rate of change of $|\beta_k\big|^2$ (with respect to 
  $\alpha h_0$) for two different regimes as follows:\\
  
 In the regime $\frac{\alpha h_0}{(k^2 + m^2)} \ll 1$, we get
 \begin{eqnarray}
 \frac{d\big|\beta_k\big|^2}{d(\alpha h_0)}&=&-\frac{1}{(k^2 + m^2)} 
 \int_0^{\infty}d\eta\cos{(2\omega\eta)} \bigg[\frac{1}{a^4(\infty)} - \frac{1}{a^4(\eta)}\bigg]\nonumber\\ 
 &\bigg[&m^2 \int_0^{\infty}d\eta\cos{(2\omega\eta)} \big[a^2(\infty) - a^2(\eta)\big] 
 - (\xi - 1/6) \int_0^{\infty}d\eta\cos{(2\omega\eta)} R(\eta)a^2(\eta)\bigg] > 0,
 \label{production3b}
\end{eqnarray}

Thereby the particle number density increases with KR field energy density for small value of $\alpha h_0$. 
Similarly for $\frac{\alpha h_0}{(k^2 + m^2)} \gg 1$,

\begin{eqnarray}
\frac{d\big|\beta_k\big|^2}{d(\alpha h_0)}&=&-\bigg[\int_0^{\infty}d\eta\cos{(2\omega\eta)} 
\bigg(\frac{1}{a^4(\infty)} - \frac{1}{a^4(\eta)}\bigg)\bigg]^2\nonumber\\
&-&2\sqrt{\alpha h_0}\bigg[\int_0^{\infty}d\eta\cos{(2\omega\eta)} \bigg(\frac{1}{a^4(\infty)} - \frac{1}{a^4(\eta)}\bigg)\bigg]
\bigg[\int_0^{\infty}d\eta~\eta \sin{(2\omega\eta)} \bigg(\frac{1}{a^4(\infty)} - \frac{1}{a^4(\eta)}\bigg)\bigg] < 0,
\label{production3a}
\end{eqnarray}
 
which indicates that $<n_k>_{out}$ decreases with KR field energy density for large $\alpha h_0$.\\
Thus as a whole, $<n_k>_{out}$ ($= \big|\beta_k\big|^2$) increases with $h_0$ for $\frac{\alpha h_0}{(k^2 + m^2)} \ll 1$ while it decreases in the regime 
$\frac{\alpha h_0}{(k^2 + m^2)} \gg 1$. This entails that $<n_k>_{out}$ must has a maximum in between these two limits of $h_0$ in a symmetric bounce 
universe, which is also reflected through 
Figure[\ref{plot1}] as the particular form of the scale factor we consider in eqn.(\ref{scale factor}) actually 
corresponds to a non-singular symmetric bounce. 
For $\frac{\alpha h_0}{(k^2 + m^2)} \ll 1$, the energy of the scalar particle can be approximated as 
$\omega_k \simeq \sqrt{k^2 + m^2}$ i.e independent of $h_0$. Now due to the coupling $f(\Phi)$, the energy supplied from the KR field to the scalar 
field increases with increasing $h_0$. This along with the fact that the energy of each scalar particle is independent of $h_0$ explains 
why the particle production is enhanced as the KR field energy density increases for $\frac{\alpha h_0}{(k^2 + m^2)} \ll 1$. On the other hand, 
for $\frac{\alpha h_0}{(k^2 + m^2)} \gg 1$, the energy of scalar particle is proportional to $h_0$, in particular $\omega_k \propto \sqrt{h_0}$. Therefore 
it becomes more difficult to excite the scalar particle and consequently the particle production decreases with increasing $h_0$ 
for $\frac{\alpha h_0}{(k^2 + m^2)} \gg 1$. These explain the inequalities obtained in eqns.(\ref{production3b}) and (\ref{production3a}) respectively.

\end{itemize}

\subsection{Quantum entanglement between scalar particles}

The ''in`` and ''out`` eigenstates of the scalar field Hamiltonian can serve as two different basis sets in the Fock space and thus the field state 
can be expressed by two ways: either by in-eigenstates or by out-eigenstates i.e

\begin{eqnarray}
 | state~\textgreater = | 0_{in}\textgreater = \sum c_n |n~\textgreater_{\vec{k}}^{out} |n~\textgreater_{-\vec{k}}^{out}
 \label{entanglement 1}
\end{eqnarray}

where $| 0_{in}\textgreater$ and $|n~\textgreater_{\vec{k}}^{out} |n~\textgreater_{-\vec{k}}^{out}$ belong from the same Fock space and also 
$c_n$ denote the corresponding coefficients between these two states. 
Since we are working in the Heisenberg picture, the field state is taken as independent of time. 
Recall, the scalar field Hamiltonian density explicitly depends on time, not space (see eqn.(\ref{hamiltonian})), 
which confirms that the energy of the scalar field is not 
a conserved quantity, but the three momentum is. This is clearly reflected through the above expression. However eqn.(\ref{entanglement 1}) further 
indicates that the field state is not separable with respect to out modes, which tells that the out modes get quantum entangled to each other. Such 
entanglement may be quantified by von-Neumann entropy ($S$) defined through the conception of reduced density operator as follows:

\begin{eqnarray}
S = -Tr\bigg[\hat{\rho}_{\vec{k}}^{red}~\log_2(\hat{\rho}_{\vec{k}}^{red})\bigg] 
\label{entanglement 2}
\end{eqnarray}
in $k_B = 1$ (Boltzmann constant) unit, 
where $\hat{\rho}_{\vec{k}}^{red}$ is the reduced density operator for $\vec{k}$th out modes and has the following definition

\begin{eqnarray}
 \hat{\rho}_{\vec{k}}^{red} = \sum \textless~m|~\hat{\rho}~|m~\textgreater_{\vec{-k}}^{out}
 \label{entanglement 3}
\end{eqnarray}

with $\hat{\rho}$ be the full density operator of the scalar field and given by
\begin{eqnarray}
 \hat{\rho} = | state~\textgreater \textless~state| = | 0_{in}\textgreater \textless~0_{in}|
 \label{entanglement 4}
\end{eqnarray}

\begin{figure}[!h]
\begin{center}
 \centering
 \includegraphics[width=3.0in,height=2.0in]{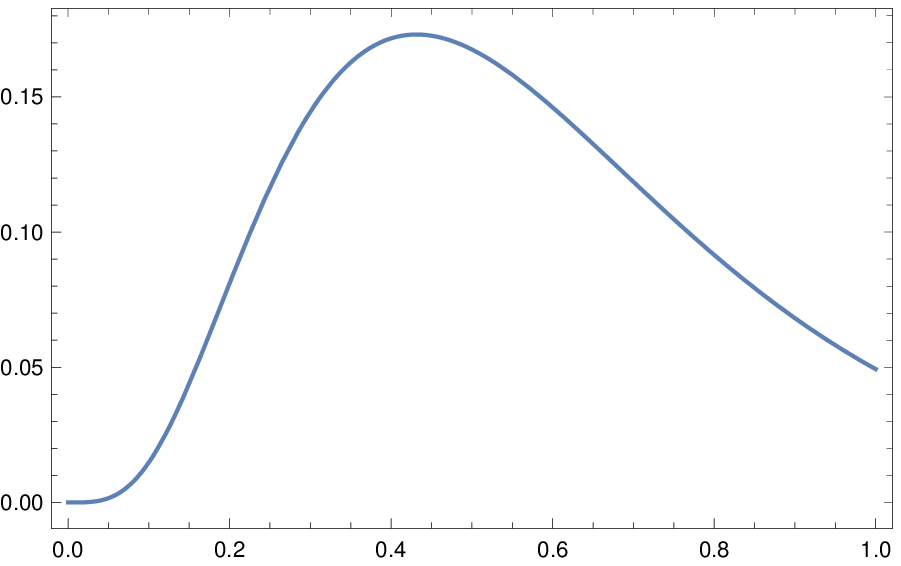}
 \includegraphics[width=3.0in,height=2.0in]{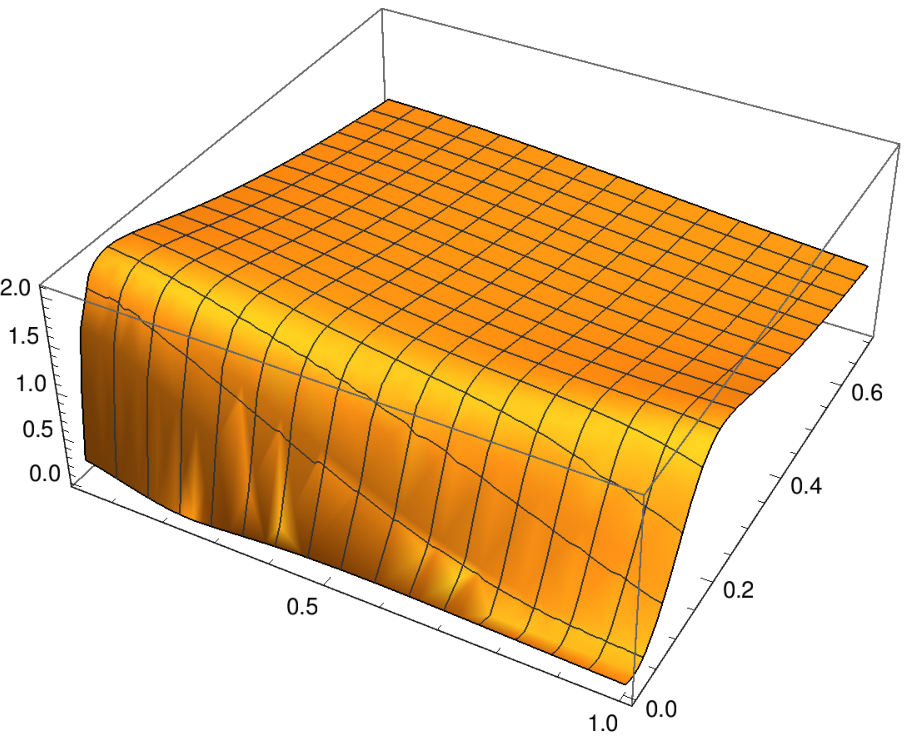}
 \caption{$Left~part$ : $S$ (along y axis) vs. $m$ (along x axis) for $\xi = 1/6$ ; $k = 0.01$ in absence of KR field. $Right~part$ : 
 3D plot of $S$ with respect to mass ($0 \leq m \leq 1$ along x axis, in Planckian unit) 
and KR field energy density ($0 \leq h_0 \leq 0.7$ along y axis for $\xi = 1/6$ ; $k = 0.01$.}
 \label{plot2}
\end{center}
\end{figure}

\begin{figure}[!h]
\begin{center}
 \centering
 \includegraphics[width=3.0in,height=2.0in]{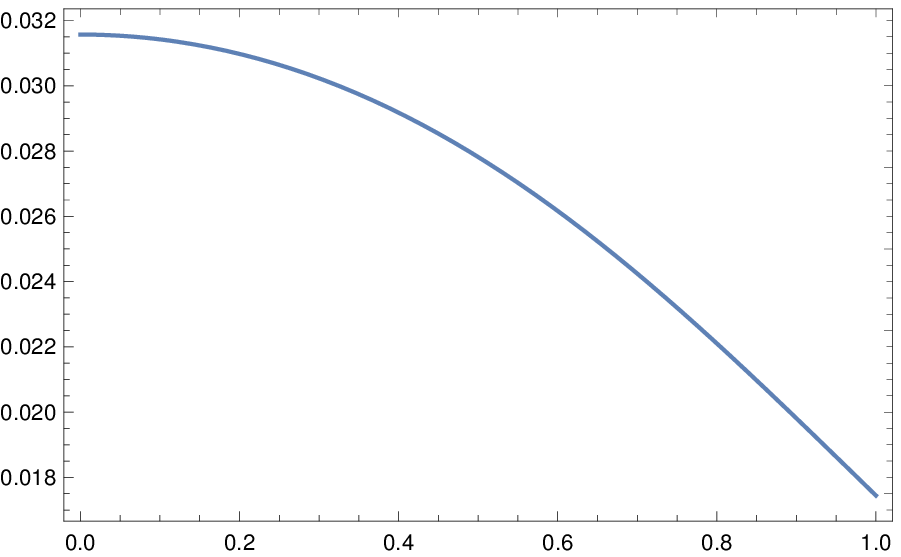}
 \includegraphics[width=3.0in,height=2.0in]{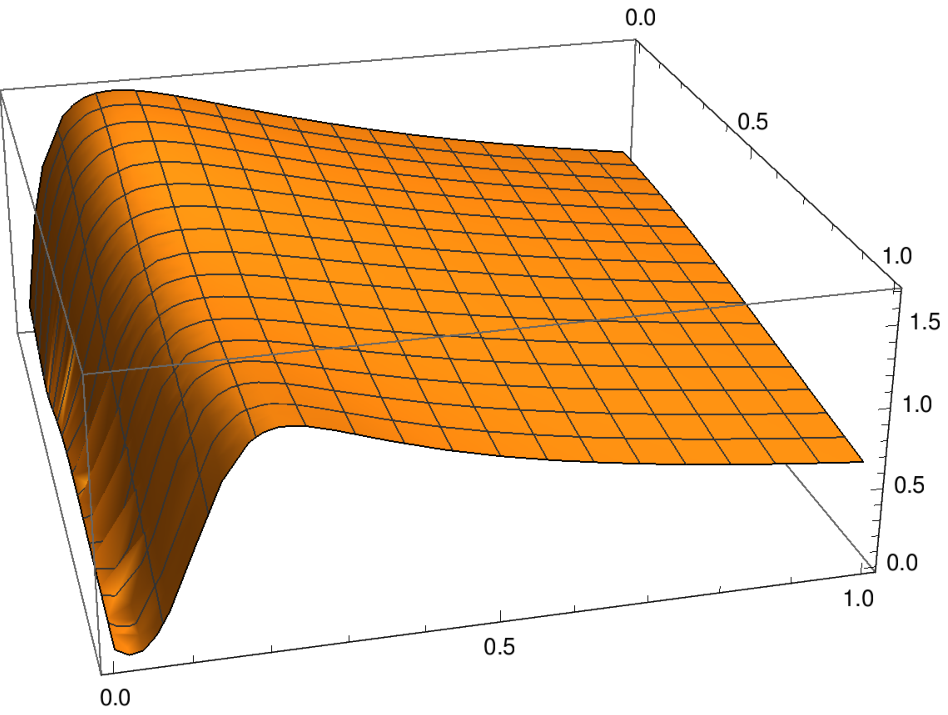}
 \caption{$Left~part$ : $S$ (along y axis) vs. $m$ (along x axis) for $\xi = 0$ ; $k = 0.01$ in absence of KR field. $Right~part$ : 
 3D plot of $S$ with respect to mass ($0 \leq m \leq 1$ along x axis, in Planckian unit) 
and KR field energy density ($0 \leq h_0 \leq 0.7$ along y axis for $\xi = 0$ ; $k = 0.01$.}
 \label{plot3}
\end{center}
\end{figure}

Going through the calculations shown in Appendix, one obtains the von-Neumann entropy in terms of Bogolyubov coefficients as follows:

\begin{eqnarray}
 S = \log_2\bigg[\frac{\gamma^{\gamma/(\gamma - 1)}}{(1 - \gamma)}\bigg]
 \label{entanglement 5}
\end{eqnarray}
with $\gamma = \bigg| \frac{\beta_k}{\alpha_k} \bigg|^2$ and the expressions of $|\alpha_k|^2$, $|\beta_k|^2$ are given in eqns.(\ref{bogolyubov 6d}), 
(\ref{bogolyubov 7d}) respectively.\\
To understand the effects of the KR field on cosmological entanglement entropy, we give the following plots : (1) Left part of Figure[\ref{plot2}] is 
the variation of entropy ($S$) with respect to mass ($m$) of the scalar field for $\xi = 1/6$ (i.e for conformal coupling) in absence of KR field, 
(2) Right part of 
Figure[\ref{plot2}] is the 3D plot exploring the variation of $S$ with respect to mass ($0 \leq m \leq 1$ along x axis, in Planckian unit) 
and KR field energy density ($0 \leq h_0 \leq 0.7$ along y axis, in Planckian unit) for $\xi = 1/6$, (3) Left and right parts of Figure[\ref{plot3}] 
give the same plots respectively for $\xi = 0$ i.e for weak coupling case.

Figure[\ref{plot2}] clearly demonstrates that in the case of conformal coupling and without the KR field, 
the entanglement entropy is bounded by $S \lesssim 0.17$ (in 
$k_B = 1$, see the left part), while in presence of KR field the upper bound of 
entropy goes beyond $0.17k_B$ and reach up to $S \lesssim 2k_B$ (see the 
right part). Therefore if the entanglement entropy is found to lie within $0.17k_B \lesssim S \lesssim 2k_B$, then it may provide a possible testbed for 
the existence of Kalb-Ramond field in our universe. Similarly Figure[\ref{plot3}] reveals that for $\xi = 0$, 
if the von-Neumann entropy situates in between $0.032k_B$ and $1.5k_B$, then one can infer about the possible presence of KR field. 
Moreover it may be noticed from Fig.[\ref{plot2}] 
that the presence of KR field causes a non-zero value of entropy even for $m = 0$ and $\xi = 1/6$ in four dimensional context. 
However this is one of the consequences of the fact 
that the presence of KR field breaks the conformal symmetry of a massless scalar field propagating in a four dimensional 
FRW spacetime (as mentioned earlier).\\
Thus it is clear that our argument regarding the testbed of Kalb-Ramond field through cosmological von-Neumann entropy relies on the fact that 
the entanglement entropy has a maxima at a finite value of $h_0$. However in the present paper we show such ''maximum`` character of entropy, but 
for a specific form of the scale factor considered in eqn.(\ref{scale factor}). Therefore it is important to investigate whether the von-Neumann 
entropy, in presence of KR field, possesses a maxima for a general class of scale factor. We investigate 
this in a bouncing universe, by determining the rate of change of $S$ with respect to $\alpha h_0$ in the regimes $\frac{\alpha h_0}{(k^2 + m^2)} \ll 1$ 
(i.e for low KR field energy density) and $\frac{\alpha h_0}{(k^2 + m^2)} \gg 1$ (i.e for large KR field energy density) respectively. Using 
eqn.(\ref{bogolyubov 5a}), we determine the expression of $\big|\alpha_k\big|^2$ for the class of symmetric bouncing scale factor as follows,

\begin{eqnarray}
 \big|\alpha_k\big|^2 = 1 + \bigg[J_1 - J_2 + J_3\bigg]^2
 \label{max_new1}
\end{eqnarray}

where the $J_i$ ($i=1,2,3$) have the following expressions:
\begin{eqnarray}
 J_1&=&m^2 \int_{-\infty}^{\infty}d\eta \big[a^2(\infty) - a^2(\eta)\big]\nonumber\\
   J_2&=&(\xi - 1/6) \int_{-\infty}^{\infty}d\eta R(\eta)a^2(\eta)\nonumber\\
   J_3&=&\alpha h_0 \int_{-\infty}^{\infty}d\eta \bigg[\frac{1}{a^4(\infty)} - \frac{1}{a^4(\eta)}\bigg]
   \label{max_new2}
\end{eqnarray}

Due to the fact $a(\eta) \leq a(\infty)$, the above integrals satisfy the inequalities as $J_1 > 0$, $J_2 \leq 0$ 
(recall we are interested on the 
weak coupling and the conformal coupling cases i.e $\xi = 0$ and $\xi = 1/6$ respectively. For $\xi = 0$, $J_2$ becomes negative, while for 
$\xi = 1/6$, $J_2$ becomes zero; that is why, as a whole we consider $J_2 \leq 0$ where the equality sign 
is for the conformal coupling) and $J_3 < 0$. Using these expressions, we obtain $\frac{d\big|\alpha_k\big|^2}{d(\alpha h_0)}$ in the 
limit $\frac{\alpha h_0}{(k^2 + m^2)} \ll 1$ as follows:

\begin{eqnarray}
 \frac{d\big|\alpha_k\big|^2}{d(\alpha h_0)}&=&\frac{1}{2(k^2 + m^2)} 
 \int_{-\infty}^{\infty}d\eta \bigg(\frac{1}{a^4(\infty)} - \frac{1}{a^4(\eta)}\bigg)\nonumber\\
 &\bigg[&m^2 \int_{-\infty}^{\infty}d\eta \big[a^2(\infty) - a^2(\eta)\big] - (\xi - 1/6) \int_{-\infty}^{\infty}d\eta R(\eta)a^2(\eta)\bigg] < 0
 \label{max_new3}
\end{eqnarray}

The first integral in the R.H.S of eqn.(\ref{max_new3}) is negative while the quantity within the square braket is positive, this results that 
$\big|\alpha_k\big|^2$ decreases with $\alpha h_0$ for small value of $\frac{\alpha h_0}{k^2+m^2}$. 
Similarly in the regime $\frac{\alpha h_0}{(k^2 + m^2)} \gg 1$ we get,

\begin{eqnarray}
 \frac{d\big|\alpha_k\big|^2}{d(\alpha h_0)} = 
 \frac{1}{4}\bigg[\int_{-\infty}^{\infty}d\eta \bigg(\frac{1}{a^4(\infty)} - \frac{1}{a^4(\eta)}\bigg)\bigg]^2 > 0
 \label{max_new4}
\end{eqnarray}

The expression $\gamma = \bigg| \frac{\beta_k}{\alpha_k} \bigg|^2$ leads to the variation of $\gamma$ with the KR field energy density as follows:
\begin{eqnarray}
 \frac{d\gamma}{d(\alpha h_0)} = \bigg| \frac{\beta_k}{\alpha_k} \bigg|^2 
 \bigg[\frac{1}{\big|\beta_k\big|^2}\frac{d\big|\beta_k\big|^2}{d(\alpha h_0)} 
 - \frac{1}{\big|\alpha_k\big|^2}\frac{d\big|\alpha_k\big|^2}{d(\alpha h_0)}
 \label{max_new5}
\end{eqnarray}

Having the above expression in hand along with the help of eqns.(\ref{production3b}), (\ref{production3a}), (\ref{max_new3}), (\ref{max_new4}), we can 
argue that  $\frac{d\gamma}{d(\alpha h_0)}$ takes positive and negative values in the limits $\frac{\alpha h_0}{(k^2 + m^2)} \ll 1$ 
and $\frac{\alpha h_0}{(k^2 + m^2)} \gg 1$ respectively. Therefore the quantity $\gamma$ must has a maxima at a finite value of $h_0$ in between these two 
limits. Eqn.(\ref{entanglement 5}) reveals that the entropy ($S$) is a monotonic increasing function of $\gamma$ by which 
we can argue that the von-Neumann entropy also possesses a maxima at a finite value of $h_0$. Hence the ''maximum`` 
character of cosmological entanglement entropy is not only confined within the scale factor as considered in eqn.(\ref{scale factor}) but also valid 
for a general class of the scale factor which corresponds to a symmetric bounce universe.\\

Before concluding, we would like to mention that the present investigation on cosmological entanglement in presence of Kalb-Ramond (KR) field 
can be extended to higher dimensions (see \cite{risi,lahanas,tuan1,tuan2,ssg_prl} for some interesting papers 
working on high dimensional scenarios of the Kalb-Ramond theory). In higher dimensional picture, for example, in a five dimensional braneworld scenario 
(two brane model), the Kalb-Ramond field is generally considered to propagate in the five dimensional bulk. Moreover on projecting 
the bulk gravity on the brane, the extra dimensional modulus field appears as a scalar field (known as radion field) in the four dimensional 
effective theory of our visible brane \cite{GW}. Thereby in the on-brane effective theory, the radion field gets coupled to the KR field \cite{tp1}. 
Thus the coupling between the scalar field and the KR field appears naturally in the five dimensional braneworld scenario, unlike to the case of four 
dimensional model where the scalar field has to be taken by hand (as we have considered the scalar field $\Phi$ in the action (\ref{action1})). 
It seems that the investigation of cosmological entanglement in presence of KR field will be interesting and is expected to be studied in near future.\\

\section{Conclusion}

In the present paper, we deal with quantum evolution of a massive scalar field propagating in a four dimensional FRW spacetime where the scalar field 
has a non-minimal coupling with the Ricci scalar. The scalar field is also coupled with a second rank antisymmetric tensor field known as 
Kalb-Ramond (KR) field. In such a scenario, we try to address the possible effects of KR field on scalar particle production and quantum entanglement 
between the produced particles with a hope that the entanglement entropy may provide a possible testbed for the existence of the KR field in our universe. 
For this purpose, the spacetime and the KR field are considered as classical fields while the scalar field is treated 
as a quantum one. The classical evolution of KR field energy density is found to decrease with the cosmological expansion of our universe 
as $1/a^6$ (with $a$ being the scale factor) i.e with a faster rate in comparison to radiation ($ \propto 1/a^4$) and matter ($\propto 1/a^3$) 
energy density. With this classical evolution, the possible effects of KR field are as follows :

\begin{enumerate}
 \item In the absence of KR field, a massless scalar field possesses conformal symmetry for $\xi = 1/6$ in four dimensional context. As a consequence, 
 there occurs no particle production and the entanglement entropy vanishes for $m = 0$ and $\xi = 1/6$ in a background FRW spacetime. 
 However the presence of KR field (actually the coupling between scalar and KR field) spoils such conformal symmetry, which in turn causes a 
 definite particle production and also a non-zero value of entanglement entropy even for $m = 0$, $\xi = 1/6$ in a 4D FRW spacetime. 
 These may be noticed from the Figures[\ref{plot1},\ref{plot2},\ref{plot3}].
 
 \item The interaction between the scalar field and the background time dependent gravitational field (i.e FRW spacetime) makes the scalar field 
 Hamiltonian explicitly time dependent which in turn excites a certain number of scalar particles ($<n_k>_{out}$) in the ''out-region`` from an 
 infinite-past-vacuum-state. Needless to say, $<n_k>_{out}$ depends on KR field energy density $h_0$ and also has a maximum at a finite value of 
 $h_0$, irrespective of conformal or weak coupling, see Fig.[\ref{plot1}]. The reason for acquiring such a maxima is that 
 for $\frac{\alpha h_0}{(k^2 + m^2)} \ll 1$, the energy of the scalar particle can be approximated as 
$\omega_k \simeq \sqrt{k^2 + m^2}$ i.e independent of $h_0$. Now due to the coupling $f(\Phi)$, the energy supplied from the KR field to the scalar 
field increases with increasing $h_0$. This along with the fact that the energy of each scalar particle is independent of $h_0$ explains 
why the particle production enhances as the KR field energy density increases for $\frac{\alpha h_0}{(k^2 + m^2)} \ll 1$. On the other hand, 
for $\frac{\alpha h_0}{(k^2 + m^2)} \gg 1$, the energy of scalar particle is proportional to $\sqrt{h_0}$, in particular 
$\omega_k \propto \sqrt{h_0}$. Therefore 
it becomes more difficult to excite the scalar particle and consequently the particle production decreases with increasing $h_0$ 
for $\frac{\alpha h_0}{(k^2 + m^2)} \gg 1$. Thus as a whole, $\frac{d<n_k>_{out}}{d(\alpha h_0)} > 0$ 
for $\frac{\alpha h_0}{(k^2 + m^2)} \ll 1$ while we get $\frac{d<n_k>_{out}}{d(\alpha h_0)} < 0$ in the regime 
$\frac{\alpha h_0}{(k^2 + m^2)} \gg 1$, which entails that $<n_k>_{out}$ must has a maximum in between these two limits.
 
 \item The scalar field evolves from an asymptotic past vacuum state to a quantum entangled state with respect to ''out-modes``. The entanglement 
 is quantified by von-Neumann entropy ($S$) in the present context. For a non-singular symmetric universe, irrespective of 
 conformal and weak coupling case, we get 
 $max[S(k, m, h_0 \neq 0)] > max[S(k, m, h_0 = 0)]$ i.e the presence of KR field makes the upper bound of the entropy larger in comparison to the case 
 when the KR field is absent. Therefore if the entanglement entropy is found to lie within these two upper bounds ( i.e within 
 $max[S(k, m, h_0 \neq 0)]$ and $max[S(k, m, h_0 = 0)]$ ), then it may provide a possible testbed for the existence of Kalb-Ramond field in our universe.\\
 
 Thereby the presence of KR field in a FRW bouncing universe allows 
 a greater particle production and consequently the upper bound of the entanglement entropy becomes larger in comparison 
 to the case when the KR field is absent. This in turn may provide a possible testbed for the existence of Kalb-Ramond field. However 
 the measurement procedure of the entanglement entropy is still a problem. If the actual method of measurement of entanglement entropy in a 
 cosmological background takes a shape in future, these predictions will certainly be useful to test the existence of a Kalb-Rammond field.
 \end{enumerate}

\section{Appendix: Detailed calculations of von-Neumann entropy}
The in-vacuum can be expressed as linear combination of out-states as 
$| 0_{in}\textgreater = \sum c_n |n~\textgreater_{\vec{k}}^{out} |n~\textgreater_{-\vec{k}}^{out}$. The coefficients $c_n$ play the most crucial 
role in determining the entanglement entropy.

\subsection{Determination of $c_n$}
As mentioned earlier, the in-vacuum is annihilated by $\hat{a}_{\vec{k}}$ i.e $\hat{a}_{\vec{k}}| 0_{in} \textgreater = 0$. Using Bogolyubov coefficients, 
one can write this annihilation condition as follows:
\begin{eqnarray}
 \big(\alpha_k^*\hat{b}_{\vec{k}} - \beta_k^*\hat{b}_{-\vec{k}}^+\big)| 0_{in} \textgreater&=&0\nonumber\\
 \Rightarrow \big(\alpha_k^*\hat{b}_{\vec{k}} - \beta_k^*\hat{b}_{-\vec{k}}^+\big)\sum c_n |n~\textgreater_{\vec{k}}^{out} 
 |n~\textgreater_{-\vec{k}}^{out}&=&0\nonumber\\
 \Rightarrow \sum c_n \alpha_k^* |n-1~\textgreater_{\vec{k}}^{out} |n~\textgreater_{-\vec{k}}^{out}&=&
 \sum c_n \beta_k^* |n~\textgreater_{\vec{k}}^{out} |n+1~\textgreater_{-\vec{k}}^{out}\nonumber\\
 \Rightarrow \alpha_k^* \sum c_{m+1} |m~\textgreater_{\vec{k}}^{out} |m+1~\textgreater_{-\vec{k}}^{out}&=&
 \beta_k^* \sum c_n |n~\textgreater_{\vec{k}}^{out} |n+1~\textgreater_{-\vec{k}}^{out}
 \label{appendix 1}
\end{eqnarray}

Eqn.(\ref{appendix 1}) clearly gives the recursion relation of the coefficients as 
\begin{eqnarray}
 c_1&=&\frac{\beta_k^*}{\alpha_k^*}c_0\nonumber\\
 c_2&=&\frac{\beta_k^*}{\alpha_k^*}c_1 = \bigg(\frac{\beta_k^*}{\alpha_k^*}\bigg)^2c_0\nonumber\\
 c_n&=&\frac{\beta_k^*}{\alpha_k^*}c_{n-1} = \bigg(\frac{\beta_k^*}{\alpha_k^*}\bigg)^2c_{n-2} 
 = ...........= \bigg(\frac{\beta_k^*}{\alpha_k^*}\bigg)^nc_0
 \label{appendix 2}
\end{eqnarray}

Therefore all the coefficients $c_n$ ($n \geq 1$) depend on the single one $c_0$ which can be determined from the normalization condition 
of $| 0_{in} \textgreater$ as follows:

\begin{eqnarray}
 \textless~0_{in}|0_{in}~\textgreater&=&1\nonumber\\
 \Rightarrow \sum |c_n|^2&=&1\nonumber\\
\Rightarrow |c_0|^2\bigg(1 + \bigg| \frac{\beta_k^*}{\alpha_k^*} \bigg|^2 + \bigg| \frac{\beta_k^*}{\alpha_k^*} \bigg|^4 + .......\bigg)&=&1\nonumber\\
\Rightarrow |c_0|^2 \bigg(\frac{1}{1 - \bigg| \frac{\beta_k^*}{\alpha_k^*} \bigg|^2}\bigg)&=&1 
\Rightarrow |c_0| = \sqrt{1 - \bigg| \frac{\beta_k^*}{\alpha_k^*} \bigg|^2} = \sqrt{1 - \gamma}
\label{appendix 3}
\end{eqnarray}

where $\gamma = \bigg| \frac{\beta_k^*}{\alpha_k^*} \bigg|^2$. Thus eqns.(\ref{appendix 2}) and (\ref{appendix 3}) lead to 
the coefficient $c_n$ (in terms of Bogolyubov coefficients) as 
\begin{eqnarray}
 c_n = \bigg(\frac{\beta_k^*}{\alpha_k^*}\bigg)^n\sqrt{1 - \gamma}
 \label{appendix 4}
\end{eqnarray}

from which the in-vacuum can be expressed (in terms of out-states) as,

\begin{eqnarray}
 | 0_{in}\textgreater = \sum \bigg(\frac{\beta_k^*}{\alpha_k^*}\bigg)^n\sqrt{1 - \gamma} 
 |n~\textgreater_{\vec{k}}^{out} |n~\textgreater_{-\vec{k}}^{out}
 \label{appendix 5}
\end{eqnarray}

\subsection{Determination of von-Neumann entropy}
The full density operator of the system is given by,

\begin{eqnarray}
 \hat{\rho}&=&| 0_{in}\textgreater \textless~0_{in}|\nonumber\\
 &=&\sum_n \sum_r 
 c_n c_r^* |n~\textgreater_{-\vec{k}}~|n~\textgreater_{\vec{k}}~\textless~r_{\vec{k}}|~\textless~r_{-\vec{k}}|
 \label{appendix 6}
\end{eqnarray}

where we use the expansion of $| 0_{in}\textgreater$ in terms of out-states. With the help of eqn.(\ref{appendix 6}), we determine the reduced density 
operator as follows :

\begin{eqnarray}
 \hat{\rho}_{\vec{k}}^{red}&=&\sum_m \textless~m_{-\vec{k}}|\hat{\rho}|m_{-\vec{k}}~\textgreater\nonumber\\
 &=&\sum_m \sum_n \sum_r c_n c_r^* \textless~m|n~\textgreater~|n~\textgreater_{\vec{k}}~\textless~r_{\vec{k}}| \textless~r|m~\textgreater\nonumber\\
 &=&\sum_m \sum_n \sum_r c_n c_r^* |n~\textgreater_{\vec{k}}~\textless~r_{\vec{k}}| \delta_{mn} \delta_{mr}\nonumber\\
 &=&\sum_n \sum_r c_n c_r^* |n~\textgreater_{\vec{k}}~\textless~r_{\vec{k}}| \delta_{nr}\nonumber\\
 &=&\sum_n |c_n|^2 |n~\textgreater_{\vec{k}}~\textless~n_{\vec{k}}|
 \label{appendix 7}
\end{eqnarray}

The entanglement entropy between the out modes is measured by von-Neumann entropy defined by,

\begin{eqnarray}
 S&=&-Tr\bigg[\hat{\rho}_{\vec{k}}^{red}~\log_2(\hat{\rho}_{\vec{k}}^{red})\bigg]\nonumber\\
 &=&- \sum_m \textless~m_{\vec{k}}|\hat{\rho}_{\vec{k}}^{red}~\log_2(\hat{\rho}_{\vec{k}}^{red})|m_{\vec{k}}~\textgreater
 \label{appendix 8}
\end{eqnarray}

Using eqn.(\ref{appendix 7}), the above expression of $S$ can be simplified as follows :

\begin{eqnarray}
 S&=&- \sum_m \textless~m_{\vec{k}}|\hat{\rho}_{\vec{k}}^{red}~\log_2(\hat{\rho}_{\vec{k}}^{red})|m_{\vec{k}}~\textgreater\nonumber\\
 &=&- \sum_n |c_n|^2 \bigg(\log_2|c_n|^2\bigg) = -|c_0|^2 \sum_n \gamma^n \log_2\bigg(\gamma^n |c_0|^2\bigg)\nonumber\\
 &=&-|c_0|^2 \bigg[\log_2\gamma \sum_n n\gamma^n + \log_2|c_0|^2 \sum_n \gamma^n\bigg]\nonumber\\
 &=&\frac{\gamma}{\gamma - 1}\log_2\gamma - \log_2\big(1 - \gamma\big) = \log_2\bigg[\frac{\gamma^{\gamma/(\gamma - 1)}}{(1 - \gamma)}\bigg]
 \label{appendix 9}
\end{eqnarray}

where we use $|c_0|^2 = (1 - \gamma)$, see eqn.(\ref{appendix 3}). Eqn.(\ref{appendix 9}) is the final expression of von-Neumann entropy in terms of 
Bogolyubov coefficients.

\end{document}